\documentclass[reprint, showpacs, superscriptaddress,nofootinbib]{revtex4-1}
\usepackage[utf8]{inputenc}
\usepackage[english]{babel}
\usepackage{amsmath}
\usepackage{amsfonts}
\usepackage{amssymb}
\usepackage{graphicx}
\usepackage{dcolumn}
\usepackage{booktabs}

\AtBeginDocument{
\heavyrulewidth=.08em
\lightrulewidth=.05em
\cmidrulewidth=.03em
\belowrulesep=.65ex
\belowbottomsep=0pt
\aboverulesep=.4ex
\abovetopsep=0pt
\cmidrulesep=\doublerulesep
\cmidrulekern=.5em
\defaultaddspace=.5em
} 
\begin{document}
\title{Atomic structure calculations of super heavy noble element oganesson (Z=118)}
\author{B.G.C. Lackenby}
\affiliation{School of Physics, University of New South Wales,  Sydney 2052,  Australia}
\author{V.A. Dzuba}
\affiliation{School of Physics, University of New South Wales,  Sydney 2052,  Australia}
\author{V.V. Flambaum}
\affiliation{School of Physics, University of New South Wales,  Sydney 2052,  Australia}
\affiliation{Johannes Gutenberg-Universit\"at Mainz, 55099 Mainz, Germany}
\begin{abstract}
We calculate the spectrum and allowed E1 transitions of the superheavy element Og (Z=118). A combination of configuration interaction (CI) and perturbation theory (PT) is used (Dzuba \textit{et at.} Phys. Rev. A, \textbf{95}, 012503 (2017)). The spectrum of lighter analog Rn I is also calculated and compared to experiment with good agreement.
\end{abstract}

\maketitle


The super heavy element (SHE) oganesson ($Z=118$) was first synthesized in 2006 at Dubna \cite{OganessianOg2006} and has recently been officially named and recognized \cite{Karol2016}.  It is also the first SHE and not naturally occuring element in the group of noble elements (Group 18) where the ground state has completely filled electron $np$ shells. Like other SHEs  ($Z>100$) it is of great experimental and theoretical interest due to the high relativistic nature  which may result in exotic and anomalous chemical and physical properties \cite{Pershina2009, Schwerdtfeger2014}. In general, experimental study of SHEs is difficult  due to the short lifetimes and low production rates. Og is no exception, where the only confirmed isotope ($^{294}$Og) has a halflife of 0.7 ms \cite{OganessianOg2006}. The study of Og and other SHEs is of great interest due to their exotic characteristics such as the large dependence on relativistic effects and the possible existence of long-lived isotopes of heavy nuclei in the ``island of stability''. \\
\linebreak
The existence of long lived SHEs is predicted to occur when the ratio of neutrons to protons ($N/Z$) is large enough for the neutron-proton attraction to overcome  the Coulombic repulsion between protons (which scales as $Z^2$). Therefore the number of neutrons must increase faster than the number of protons requiring extremely neutron-rich isotopes to be  long-living\cite{OUL2004, HHO2013}.  Early nuclear shell models predict the nuclear shells stabilize for the ``magic'' numbers $Z=114$ and $N=184$\cite{OUL2004, HHO2013}. Synthesizing these neutron-rich isotopes is an extremely difficult challenge as the collision of two nuclei with a smaller $N/Z$ will always result in a neutron poor element.  However an alternate route to identify these long lived SHEs may be through analysing astrophysical data. Such avenues have already been explored with astrophysical data of  Przybylski's star suggesting that elements up to $Z=99$ have probably  been identified\cite{Polukhina2012, Gopka2008, Fivet2007}.  They may be decay products of long lived nuclei (see e.g.   \cite{DFW17} and references therein).  It is suspected that neutron rich isotopes may be created in cosmic events where rapid neutron capture (``$r$-process'') can occur due to large neutron fluxes during supernovae explosions, neutron star - (black hole and neutron star) mergers \cite{Goriely2011, Fuller2017, Friebel2018, Schuetrumpf2015}. To predict  atomic transition frequencies  for the neutron-reach isotopes the  calculated isotopic shifts should be added to the atomic transition frequencies measured in laboratories for the neutron-poor isotopes \cite{DFW17}. Search for these SHE in astrophysical data requires the strong electric dipole (E1) transitions which we calculate in this work.

 There has been a large amount of  theoretical work on the chemical and physical properties of Og with calculations of solid state and molecular properties \cite{Kullie2012, Shee2015, Nash1999, Nash2005, Peter2016}, electron affinities\cite{Pitzer1975, EliavOg1996, PershinaOg2008, Hangele2012, Goidenko2003}, and ionisation potentials and polarisabilities \cite{PershinaOg2008, Desclaux1973, Nash2005, Jerabek2018}. While some odd parity states and electric dipole (E1) transitions in the Og spectrum have been calculated in \cite{Indelicato2007} we present a more complete spectrum with both odd and even states to compare against similar states in the Rn spectrum. \\
There has been considerable work on both relativistic and quantum electrodynamic (QED) effects \cite{Pyykko1988, Jerabek2018, Goidenko2003, Eliav2015, Indelicato2007, Thierfelder2010} in Og. In this work we included both the Breit interaction and QED radiative effects. To aid in the experimental study of Og we use  theoretical methods to further study its physical properties. \\

\section{CIPT calculation of R\MakeLowercase{n} I and O\MakeLowercase{g} I} \label{sec:CIPT}
\begin{table*} [t!]
\begin{center}

\caption{CIPT calculations of excitation spectrum, ionisation potential and electron affinity for Rn I and Og I. Experimental results for Rn I are included for comparison. Here $E_E$ and $E_T$ are experimental and theoretical CIPT excitation energies respectively with $\Delta = E_E - E_T$. We also present the calculated Land\'{e} $g$-factors and the energy difference between the experimental and theoretical excitation energies.  \label{tab:RnOgSpectrum}}
\begin{tabular}{l@{\hspace{0.5cm}}cc@{\hspace{0.5cm}}r@{\hspace{0.5cm}}r@{\hspace{0.5cm}}r@{\hspace{0.5cm}}r@{\hspace{0.5cm}}|l@{\hspace{0.5cm}}cc@{\hspace{0.5cm}}r@{\hspace{0.5cm}}r@{\hspace{0.5cm}}r}
\toprule
\toprule
\multicolumn{7}{c}{Rn I} & \multicolumn{5}{c}{Og I} \\
 & State & $J$ &  \multicolumn{1}{c}{\parbox{1cm}{$E_E$\cite{NIST_ASD} \\ (cm$^{-1}$)}}  &  \multicolumn{1}{c}{\parbox{1cm}{$E_T$ \\ (cm$^{-1}$)}} &  \multicolumn{1}{c}{$g_T$} &  \multicolumn{1}{c}{\parbox{1cm}{$\Delta$ \\ (cm$^{-1}$)}} &  & State & $J$ & \multicolumn{1}{c}{\parbox{1cm}{$E_T$ \\ (cm$^{-1}$)}}  &  \multicolumn{1}{c}{$g_T$} & \parbox{1.5cm}{Ref. \cite{Indelicato2007}\\ (cm$^{-1}$)}  \\
 \hline
 \\
$6s^2 6p^6$      & $^1$S & 0 & 0    & 0    & 0    &      &   $7s^2 7p^6$ & $^1$S & 0  &  0 & 0 & 0  \\
$6s^2 6p^5 7s$ &    $^3$P$^{\rm_o}$         & 2  &  54 620  &  55 323     &  1.50     &  -703    & $7s^2 7p^5 8s$  &  $^3$P$^{\rm_o}$  & 2 & 33 884 & 1.50 & 34 682 \\
$6s^2 6p^5 7s$ &  $^1$P$^{\rm_o}$  & 1 &   55 989  & 56 607   &  1.18      &   -618  & $7s^2 7p^5 8s$ & $^1$P$^{\rm_o}$ &  1   & 36 689    &  1.17  & 38 150   \\
$6s^2 6p^5 7p$ & $^3$S  &  1 &   66 245 & 67 171   & 1.76      &   -926   & $7s^2 7p^5 8p$ & $^3$P &  1   & 49 186    &    1.60  &   \\
$6s^2 6p^5 7p$ & $^3$D &  2 &   66 708 & 67 658  & 1.13     &   -950   & $7s^2 7p^5 8p$ & $^3$D & 2    &  49 451   &   1.15    \\
$6s^2 6p^5 6d$ & $^1$S$^{\rm_o}$ &   0 &  67 906  &   69 145  &  0       &  -1 239    & $7s^2 7p^5 8p$& $^3$D & 3    &  53 777   &    1.33 &    \\
$6s^2 6p^5 7p$ & $^3$D &  3 &   68 039  &  68 891  &  1.33      &   -852  & $7s^2 7p^5 8p$ & $^3$P &  1   & 53 881    &    1.24        \\
$6s^2 6p^5 7p$ & $^1$P & 1 &   68 332 &  69 313   & 1.09      & -981   &  $7s^2 7p^5 7d$ &  $^1$S$^{\rm_o}$ & 0    & 54 155    &  0  & 53 556    \\
$6s^2 6p^5 7p$ & $^3$P &  2 &   68 790 &  69 749   & 1.37       &  -959   &  $7s^2 7p^5 8p$ & $^3$P & 2     & 54 446 & 1.35 &  \\
$6s^2 6p^5 6d$ & $^3$P$^{\rm_o}$ &  1 &  68 891  & 70 002     &   1.36    & -1 111  &  $7s^2 7p^5 7d$ &  $^1$S$^{\rm_o}$ & 1    & 54 725    &  1.33 & 54 927 \\
$6s^2 6p^5 7p$ & $^1$S &  0 &   69 744    &   70 800    &   0    &   -1 056   &  $7s^2 7p^5 7d$ & $^3$F$^{\rm_o}$ & 4    &  54 938   &     1.25 & 48 474     \\
$6s^2 6p^5 6d$ & $^3$F$^{\rm_o}$ &  4 &    69 798   &   70 742     &  1.25    &  -944   & $7s^2 7p^5 7d$ & $^3$D$^{\rm_o}$ & 2    &  55 416    &    1.30 & 49 039 \\
$6s^2 6p^5 6d$ &  $^3$D$^{\rm_o}$ &  2     &   70 223   &  71 188 &   1.32    &   -965   & $7s^2 7p^5 7d$ & $^3$F$^{\rm_o}$ &   3  &  55 622   &  1.06 & 49 603  \\
$6s^2 6p^5 6d$ & $^3$F$^{\rm_o}$ &  3 & 70 440  &  71 334 &  1.06     &    -894   & $7s^2 7p^5 8p$ & $^1$S & 0  &  55 729 & 0     \\
& & & & & &    & $7s^2 7p^5 7d$ & $^1$D$^{\rm_o}$ &  2   &  56 317   & 0.98 & 50 410  \\
& & & & & & & $7s^2 7p^5 7d$ & $^5$F$^{\rm_o}$ &  3   & 56 343    &   1.25 & 50 168 \\
& & & & & & & $7s^2 7p^5 7d$ & $^1$P$^{\rm_o}$ &  1   &  57 855   &   0.84  & 58 072  \\
\multicolumn{12}{c}{Ionisation potentials} \\
$6s^2 6p^5$ & $^2$P$^{\rm_o}$ & $3/2$ & 86 693 & 87 721 & 1.33 & -1 028     & $7s^2 7p^5$  & $^2$P$^{\rm_o}$ &   3/2  & 71 508    & 1.33  & 71 320\cite{Jerabek2018}    \\
\multicolumn{12}{c}{Electron Affinity} \\
$6s^2 6p^6 7s$ &   $^2$S & 1/2 &  & 1 868 & 2.00 &      & $7s^2 7p^6 8s$  & $^2$S  & 1/2    & -773$^{a}$    & 2.00  & -516 \cite{Goidenko2003}    \\

\bottomrule
\bottomrule
\end{tabular}
\end{center}
\begin{flushleft}
$^a$ Negative value indicates the state is bound.
\end{flushleft}
\end{table*}
To calculate the spectra of oganesson we use a combination of the configuration interaction and perturbation theory (CIPT), introduced in ref. \cite{DBHF2017}. This technique has been used to calculate the spectra in open $d$-shell and open $f$-shell atoms with a large number of valence electrons where other many-body methods are unfeasible\cite{DBHF2017, LDFDb2018, Dzuba2018}. Calculations for W I, Ta I and Yb I are in good agreement with experiment. In this section we will give a brief overview of the CIPT method for Rn and Og. For an in depth discussion of the CIPT method refer to refs. \cite{DBHF2017}. \\

We generate the set of complete orthogonal single-electron states for both Rn I and Og I by using the $V^{N-1}$ approximation \cite{Kelly1964, Dzuba2005} (where $N$ is the total number of electrons). The Hartree-Fock (HF) calculations for atomic core are done for the open-shell configurations $6s^2 6p^5$ and $7s^2 7p^5$ for the Rn I and Og I respectively. The single-electron basis sets are calculated in the field of the frozen core using a B-spline technique with 40 B-spline states of order 9 in a box with radius 40 $a_B$ (where $a_B$ is the Bohr radius) with partial waves up to $l_{\text{max}} = 4$ included \cite{Johnson1988}. \\

The many-electron wavefunctions $ |i \rangle = \Phi_i(r_1,\dots,r_{N_e})$ are formed through single and double excitations from low-lying reference configurations. The many-electron wavefunctions are ordered by energy and divided into two sets. The first set represents a small number of low energy states which contribute greatly to the total CI valence wavefunction ($i < N_{\text{eff}}$, where $N_{\text{eff}}$ is the number of included low energy states) and the remaining wavefunctions represent a large number of high energy terms which are small corrections to the valence wavefunction ($N_{\rm eff} < i \le N_{\text{total}}$). The valence wavefunction can be written as 
\begin{align} 
| \Psi \rangle &= 
\sum_{i=1}^{N_{\text{eff}}} c_{i}|i\rangle + \sum_{i = N_{\text{eff}} + 1}^{N_{\text{total}}} c_{i}|i\rangle .
\end{align}
The off-diagonal matrix elements between the higher order states are neglected ($\langle i | H^{\text{CI}} | j \rangle = 0 $ for $N_{\rm eff} < i,j \le N_{\text{total}}$) which greatly decreases the computation time for a small sacrifice in accuracy. \footnote{It immediately follows from the perturbation theory that contributions of CI matrix elements between high states to low state energy are suppressed by a second power of large energy denominators while the contribution of matrix elements between high and low states are only suppressed by the first power in the denominator.}  

The matrix elements between high energy and low energy states are included pertubatively by modifying the low energy matrix elements,
\begin{equation}
\langle i|H^{\rm CI}|j\rangle \rightarrow \langle i|H^{\rm CI}|j\rangle + 
\sum_k \frac{\langle i|H^{\rm CI}|k\rangle\langle k|H^{\rm
    CI}|j\rangle}{E - E_k}, 
\end{equation}
 where $i,j \le N_{\rm eff}$, $N_{\rm eff} < k \le N_{\rm total}$,  $E_k = \langle k|H^{\rm CI}|k\rangle$, and $E$ is the energy of the state of interest.  This results in a modified CI matrix and the energies  are found through solving the standard eigenvalue problem,
\begin{align} \label{eq:CI_diag}
(H^{\rm CI} - EI)X=0,
\end{align} 
where $I$ is unit matrix, the vector $X = \{c_1, \dots, c_{N_{\rm eff}}\}$. The CI equations (\ref{eq:CI_diag}) are iterated in the CIPT method. For a detailed discussion of the CIPT precedure see Refs.~\cite{DBHF2017, LDFDb2018}. \\

We included both Breit interaction\cite{Breit1929, Mann1971, DF2016}  and QED radiative corrections in our calculation of the Og spectra.  The Breit interaction $V_B$ accounts for the magnetic interaction between two electrons and retardation. The QED corrections $V_R$ accounts for the Ueling potential and electric and magnetic formfactors\cite{FG2005}. 

For the calculation of the even parity states of Og the low energy reference states in the effective matrix were $7s^27p^6$ and $7s^27p^58p$ while for the odd states $7s^2 7p^5 8s$ and $7s^2 7p^5 7d$. For the calculation of the ionisation potential and electron affinity we remove or add one electron from the states in the effective matrices respectively.\\
 
Each level is presented with an $LS$ notation. These are selected by comparing calculated $g$-factors to the non-relativistic expression,
\begin{align} 
g_{NR} =  1 + \dfrac{J(J + 1) - L(L+1) + S(S+1)}{2J(J+1)}.
\end{align}
and using the $L$ and $S$ values as fitting parameters. We stress that the presented $LS$ notations are approximations as the states of Og are highly relativistic and strongly mixed. \\

In Table~\ref{tab:RnOgSpectrum} we present the results of our CIPT calculations for Rn I and Og I. We compare the Rn I CIPT calculations to the experimental results. The lack of experimental $g$-factors for Rn I make it difficult to confirm the correct identification of the states and therefore we must rely solely on the order of the energy levels. We find that there is good agreement between the experimental and theoretical states with an agreement with  $\Delta \approx -900$ cm$^{-1}$ with the largest discrepancy   $\Delta \approx -1239 $~cm$^{-1}$. We expect a similar accuracy for our Og I calculations (also presented in Table~\ref{tab:RnOgSpectrum}). \\

Comparing the spectrum of Rn  to Og we see that despite the similar electronic structure (with differing principal quantum numbers) there are significant differences. The Og spectrum is much more dense than Rn  with the first excitation lying more than 20~000~cm$^{-1}$ below the equivalent excitation in Rn. This results in an odd parity state which lies in the optical region. This makes the  state a good candidate for initial experimental measurement. In the final column of Table~\ref{tab:RnOgSpectrum} we present the states calculated in ref.~\cite{Indelicato2007}. This work also did not present $g$-factors which made comparing states uncertain, therefore we compared them by ordering energies. For 4 of the states there was good agreement with our results lying within $1000$ cm$^{-1}$ however for the other states there was a large discrepancy of $>4000$ cm$^{-1}$.  \\

Our calculated value of the ionisation potential of Og in Table~\ref{tab:RnOgSpectrum} is in excellent agreement with the value calculated in Ref.~\cite{Jerabek2018} ($E_{IP}=$71~320~cm$^{-1}$) where a CCSD(T) method was used.  

It has been shown that Og has a positive electron affinity which is an anomaly in the group of noble gases \cite{EliavOg1996, Goidenko2003, Eliav2015}. This is another consequence of the stabilized $8s$ orbital due to the large relativistic effects. Our calculation presented in Table \ref{tab:RnOgSpectrum}  confirms this with an electron affinity of 773 cm$^{-1}$ (0.095 eV) which is in good agreement with the coupled cluster value presented in \cite{Goidenko2003}. For comparison we also present the negative ion calculation for Rn I which is known to be unstable. All other negative ionic states of Og were found to be unstable.

\section{Electric dipole transitions of O\MakeLowercase{g} I} \label{sec:E1}
 While Og  follows the expected trend for elements in noble group where each consecutive element has both a smaller IP and first excitation energy. However Og has some properties which can be considered exotic even amongst the Group 18 elements. According to the calculated spectrum in Table~\ref{tab:RnOgSpectrum} it is the only noble element which has an allowed optical electric dipole (E1) transition ($\omega < $40~000 cm$^{-1}$) from the ground state, unlike Rn where the first odd state lies at 57~334 cm$^{-1}$. \\
 
 The E1 transition amplitudes, $A_{\text{E1}}$, between states which satisfy the conditions of opposite parity and $\Delta J \leq 1$   are calculated using the many-electron wavefunctions created in the CIPT method and the self-consistent random-phase approximation which 
includes polarization of  the atomic electron core by an external electromagnetic field. The details  of the method are presented  in Ref. \cite{Dzuba2018}. \\
 
The E1 transition rate is calculated using (in atomic units),
\begin{align} \label{eq:Transitionrate}
T_{E1} = \dfrac{4}{3}\left(\alpha \omega\right)^3\dfrac{ A_{\text{E1}}^2}{2J + 1}
\end{align}
where $J$ is the angular momentum of the upper state, $\alpha$ is the fine structure constant and $\omega$ is the frequency of the transition in atomic units. The transition amplitudes and transition rates for the allowed E1 transitions in Og are presented in Table~\ref{tab:E1_transitions}. In ref.~\cite{Indelicato2007} the major E1 transition rates were also calculated with a MCDF approach, these are included for comparison Table~\ref{tab:E1_transitions}. \\
\linebreak
We calculated the  rates of the $(n+1)s \rightarrow np$ transitions in lighter neutral noble elements Kr and Xe  and compared them to experimental values, these are presented in Table \ref{tab:E1_comp}. The experimental uncertainties are approximately $2\%$ for Xe I transitions\cite{Xe_BSD} and $~10$-$25\%$ for Kr I transitions\cite{Kr_BSD}. Comparing our calculated values to the experimental values in Table \ref{tab:E1_comp} we see the accuracy for these transitions is from 0.6\% to 17.7\%. We used the experimental energies to calculate the transitions rates of Kr I and Xe I using (\ref{eq:Transitionrate}) and since the uncertainty in the experimental energies are negligible the uncertainty in our calculations compared to experimental results in Table \ref{tab:E1_comp} is equivalent to the uncertainty in the square of the calculated transition amplitude $A_{E1}^2$. For our calculation of the Og I transition rates we needed to take into account the non-negligible uncertainty in the energies of our CIPT calculations.  Therefore assuming an accuracy of 18\% for $A_{E1}^2$ and an uncertainty of 3\% in the CIPT energy ($\left|\Delta\right| \approx 1000$~cm$^{-1}$) we expect a transition rate accuracy of 20\% for the $8s \rightarrow 7p$ optical transition ($\omega = 36~689$~cm$^{-1}$) of Og I in Table \ref{tab:E1_transitions}.  \\
\begin{table}[h]
\center
\caption{Comparison of E1 transition rates between experimental and CIPT values for Kr I and Xe I. Here $A_{E1}$ is the transition amplitude in atomic units and $T_{E1}$ is the transition rate. \label{tab:E1_comp}}
\begin{tabular}{c@{\hspace{0.5cm}}c@{\hspace{1cm}}c@{\hspace{0.5cm}}c@{\hspace{0.5cm}}c}
\toprule
\toprule
State & $E_{\text{Exp}}$ & $A_{\text{E1}}$ & $T_{\text{E1, CIPT}}$ & $T_{\text{E1,Exp}}$  \\
&  (cm$^{-1}$) & (a.u.) &  ($\times 10^6$ s$^{-1}$) &  ($\times 10^6$ s$^{-1}$)  \\
\hline
\multicolumn{5}{c}{Kr I} \\
$^1$P$_1^{\rm_o}$ & 80 916 & 0.94  & 314  & 312\cite{Kr_BSD} \\
 $^3$P$_1^{\rm_o}$ & 85 846 & 0.87  & 320  & 316\cite{Kr_BSD}   \\
 \multicolumn{5}{c}{Xe I} \\
 $^1$P$_1^{\rm_o}$ & 68 045 & 1.18  & 295  & 273\cite{Xe_BSD} \\
 $^3$P$_1^{\rm_o}$ & 77 185 & 0.98  & 298  & 253\cite{Xe_BSD}  \\
\bottomrule
\bottomrule
\end{tabular}
\end{table}

\begin{table}[h]
\center
\caption{Electric dipole transition amplitudes of Og I from the ground state $^1$S$_0$ to the excited states of odd parity and angular momenta $J=1$. Here $A_{E1}$ is the transition amplitude in atomic units and $T_{E1}$ is the transition rate. We include results of MCDF calculations from ref. \cite{Indelicato2007} for comparision. There is significant disagreement for the third transition however there is another transition in \cite{Indelicato2007} which has a rate ($986 \times 10^{6}$ s$^{-1}$) close to our calculated value. So, the disagreement  may be the result of a misprint in \cite{Indelicato2007}.\label{tab:E1_transitions}}
\begin{tabular}{c@{\hspace{0.5cm}}c@{\hspace{1cm}}c@{\hspace{0.5cm}}c@{\hspace{0.5cm}}c}
\toprule
\toprule
State & $E_{\text{CIPT}}$ & $A_{\text{E1}}$ & $T_{\text{E1, CIPT}}$ & $T_{\text{E1,MCDF}}$ \cite{Indelicato2007}  \\
&  (cm$^{-1}$) & (a.u.) &  ($\times 10^6$ s$^{-1}$) &  ($\times 10^6$ s$^{-1}$)  \\
\hline
$^1$P$_1^{\rm_o}$ & 36 689 & 2.09 & 145 & 204\\
 $^1$S$_1^{\rm_o}$ & 54 725 & 0.727 & 58.4 & 55.3  \\
 $^1$P$_1^{\rm_o}$ & 57 855 & -2.67 & 936 & 9.9, 986* \\
\bottomrule
\bottomrule
\end{tabular}
\end{table}
Only the first transition in Table \ref{tab:E1_transitions} lies in the optical region and therefore it has the highest likelihood of being measured first. The large rate of the transition $^1$S$_0$~$\rightarrow$~$^1$P$_1^{\rm_o}$ is also promising for experimental measurement. 

\section{Electron density of O\MakeLowercase{g} } \label{sec:Relativistic}
\begin{figure} 
\includegraphics[scale=0.58]{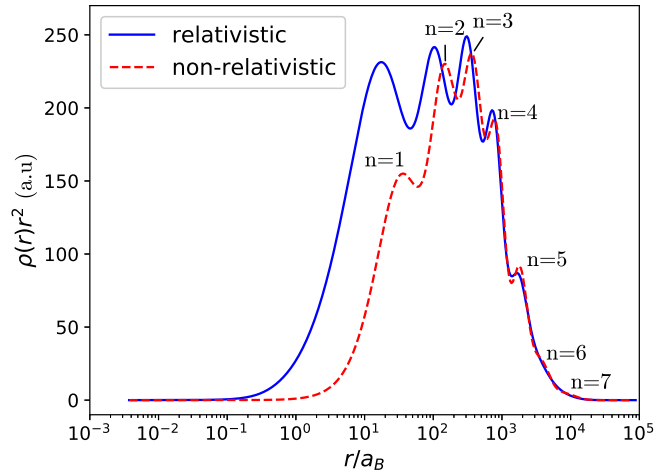}
\caption{Radial electron density, $4\pi\rho(r)r^2$ plot for Og I in both relativistic and non-relativistic approximations. The solid blue line and the dashed red line are non-relativistic and relativistic approximations respectively. The principle quantum peaks have been labeled for the non-relativic plot.\label{Og_plot}}
\end{figure}
\begin{figure}
\includegraphics[scale=0.55]{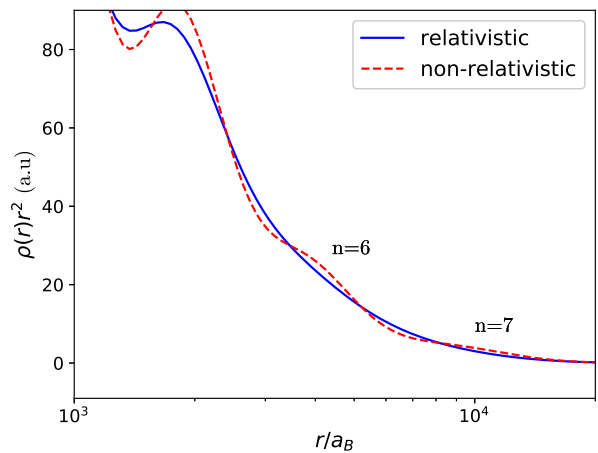}
\caption{Lower right section of  Figure~\ref{Og_plot}.\label{Og_plot_zoom}}
\end{figure}
It has been shown in Ref. \cite{Jerabek2018} using fermion localization that the electron density of Og is smoother than other group 18 analogues which have distinct atomic shells . The cause of this is the large relativistic effects in SHE which effectively smear out the shells into a smoother electron density (the same was shown for the nucleon density). The relativistic effects can also be seen by looking at the radial electron densities with relativistic and non-relativistic approximations. The Hartree-Fock radial electron density for Og is plotted on a logarithmic scale in Figure~\ref{Og_plot} in both the relativistic and non-relativistic approximations. There are a total of 7 peaks in the radial densities corresponding to the principle quantum numbers $n$ where lower shells have distinct peaks in both the relativistic and non-relativistic approximations. As expected, in the relativistic approximation the inner shells ($n=1,2,3$) shift closer to the nucleus however higher shells are relatively unaffected ($n \geq 4$). This results in a similar density profile for the electrons a large distance away from the nucleus. \\

In Figure~\ref{Og_plot_zoom} we plot the tail of the density function in Figure~\ref{Og_plot}. Here we see that, while spread out, the principle shell peaks still exist in the non-relativistic approximation. However in the relativistic approximation the density has been smoothed out to such a degree that there are no discernible peaks. This supports the results in ref.~\cite{Jerabek2018} where they calculated the electron shell structure of Og I and found that it disappears for external shells due to the high relativistic effects. This can be explained as the large spin-orbit splitting doubles the number of sub-shells which overlap making the overall distribution smooth.\\

\section{Conclusion}
In this work we calculated the spectrum and E1 transitions for Og I including the ionisation potential. We demonstrated the accuracy of the calculations by comparing similar calculations of Rn I to experimental data and expect an uncertainty of no more than $|\Delta| \approx 1000$~cm$^{-1}$. We found the spectrum of Og I is dense compared to other elements in  group 18 with significantly lower ionisation potential and excited states which follows the periodic trend. This compact spectrum introduces an allowed optical E1 transition which does not exist in other group 18 elements which presents a possibility for future experimental measurements. Our work also supports recent findings\cite{Jerabek2018} which suggest the electron shell structure of Og I is less prominent than lighter elements due to large relativistic effects which results in the outer electron density to becoming smooth. \\

This work was funded in part by the Australian Research Council.

\bibliographystyle{apsrev4-1}
\bibliography{SHE}

\end{document}